%% file: main.tex
\documentclass[letterpaper,twocolumn,10pt]{article}
\usepackage{usenix-2020-09}

\usepackage{xcolor}
\usepackage{listings}
\definecolor{codegray}{rgb}{0.1,0.1,0.1}

\definecolor{darkblue}{RGB}{20, 20, 139}
\definecolor{darkgreen}{RGB}{0, 100, 0}
\definecolor{darkred}{RGB}{139, 0, 0}
\definecolor{darkorange}{RGB}{215, 110, 0}

\definecolor{darkpurple}{RGB}{75, 0, 130}
\definecolor{darkgray}{RGB}{105, 105, 105}
\definecolor{darkcyan}{RGB}{0, 139, 139}
\definecolor{darkteal}{RGB}{0, 128, 128}

\lstdefinestyle{compact}{
    keywordstyle=\color{darkblue},
    commentstyle=\color{darkgreen},
    stringstyle=\color{darkorange},
    literate=
        *{0}{{\textcolor{darkred}{0}}}1
        {1}{{\textcolor{darkred}{1}}}1
        {2}{{\textcolor{darkred}{2}}}1
        {3}{{\textcolor{darkred}{3}}}1
        {4}{{\textcolor{darkred}{4}}}1
        {5}{{\textcolor{darkred}{5}}}1
        {6}{{\textcolor{darkred}{6}}}1
        {7}{{\textcolor{darkred}{7}}}1
        {8}{{\textcolor{darkred}{8}}}1
        {9}{{\textcolor{darkred}{9}}}1,
    numberstyle=\tiny\color{codegray},
    basicstyle=\fontsize{8}{5}\selectfont\ttfamily,
    columns=fullflexible,
    breakatwhitespace=false,
    breaklines=true,                 
    captionpos=b,                    
    keepspaces=true,                 
    numbers=left,                    
    numbersep=3pt,                  
    showspaces=false,                
    showstringspaces=false,
    showtabs=false,                  
    tabsize=2,
    frame=none, 
    framexbottommargin=0pt, 
    framextopmargin=0pt, 
}

\lstdefinestyle{compact-alt}{
    keywordstyle=\color{darkpurple},
    commentstyle=\color{darkgray},
    stringstyle=\color{darkteal},
    literate=
        *{0}{{\textcolor{darkcyan}{0}}}1
        {1}{{\textcolor{darkcyan}{1}}}1
        {2}{{\textcolor{darkcyan}{2}}}1
        {3}{{\textcolor{darkcyan}{3}}}1
        {4}{{\textcolor{darkcyan}{4}}}1
        {5}{{\textcolor{darkcyan}{5}}}1
        {6}{{\textcolor{darkcyan}{6}}}1
        {7}{{\textcolor{darkcyan}{7}}}1
        {8}{{\textcolor{darkcyan}{8}}}1
        {9}{{\textcolor{darkcyan}{9}}}1,
    numberstyle=\tiny\color{codegray},
    basicstyle=\fontsize{7.3}{5}\selectfont\ttfamily,
    columns=fullflexible,
    breakatwhitespace=false,
    breaklines=true,                 
    captionpos=b,                    
    keepspaces=true,                 
    numbers=left,                    
    numbersep=3pt,                  
    showspaces=false,                
    showstringspaces=false,
    showtabs=false,                  
    tabsize=2,
    frame=none, 
    framexbottommargin=0pt, 
    framextopmargin=0pt, 
}

\usepackage{soul}
\usepackage[font=small,skip=0pt]{caption}
\usepackage{textcomp}

\usepackage[normalem]{ulem}
\usepackage{tikz}
\usepackage{comment}
\usepackage[utf8]{inputenc}
\usepackage{graphicx}
\usepackage{subcaption}
\usepackage{fancyvrb}
\usepackage{multirow}
\usepackage{makecell}
\usepackage{array, multirow}
\usepackage{arydshln}
\newcommand{\ignore}[1]{}
\usepackage{balance}
\usepackage[ruled,vlined]{algorithm2e}
\usepackage{booktabs}
\usepackage{adjustbox}
\usepackage{placeins}
\usepackage{environ}
\usepackage[smartEllipses]{markdown}
\usepackage{float}

\usepackage{pifont}
\usepackage{wasysym}

\newcolumntype{P}[1]{>{\centering\arraybackslash}p{#1}}
\usepackage{colortbl}
\usepackage{array}
\definecolor{Gray}{gray}{0.85}
\newcolumntype{H}{@{}>{\setbox0=\hbox\bgroup}c<{\egroup}}
\newcolumntype{a}{>{\columncolor{Gray}}c}

\usepackage{epigraph}

\usepackage{tikz}
\newcommand*\circled[1]{\tikz[baseline=(char.base)]{
            \node[shape=circle,draw,inner sep=1pt,semithick] (char) {\small #1};}}

\newcommand{\del}[1]{}

\NewEnviron{delenv}{}

\usepackage[available,functional]{usenixbadges}

\usepackage{authblk}

\begin{document}

\date{}

\title{\Large \bf Attacking with Something That Does Not Exist:\\`Proof of Non-Existence' Can Exhaust DNS Resolver CPU}

\renewcommand\Authand{, }
\renewcommand\Authands{, }

\author[*\dag]{Olivia~Gruza}
\author[*\dag]{Elias~Heftrig}
\author[*\dag]{Oliver~Jacobsen}
\author[*\dag]{Haya~Schulmann}
\author[*\dag]{Niklas~Vogel}
\author[*\ddag{}$\mathsection$]{Michael~Waidner}

\affil[*]{National Research Center for Applied Cybersecurity ATHENE}
\affil[{$\dagger$}]{Goethe-Universit\"at Frankfurt}
\affil[$\ddagger$]{Technische Universit\"at Darmstadt}
\affil[$\mathsection$]{Fraunhofer Institute for Secure Information Technology SIT}

\maketitle

\input{000-abstract}

\input{010-introduction}

\input{020-background}

\input{040-attack}

\input{050-impact}

\input{060-meaurements}

\input{061-works}
\input{070-conclusions}

\section*{Acknowledgements}
This work has been co-funded by the German Federal Ministry of Education and Research and the Hessen State Ministry for Higher Education, Research and Arts within their joint support of the National Research Center for Applied Cybersecurity ATHENE and by the Deutsche Forschungsgemeinschaft (DFG, German Research Foundation) SFB~1119.

\balance

\bibliographystyle{plain}
\bibliography{main}

\balance
\appendix

\end{document}

%% file: 000-abstract.tex
\begin{abstract}
NSEC3 is a proof of non-existence in DNSSEC, which provides an authenticated assertion that a queried resource does not exist in the target domain. NSEC3 consists of alphabetically sorted hashed names before and after the queried hostname. To make dictionary attacks harder, the hash function can be applied in multiple iterations, which however also increases the load on the DNS resolver during the computation of the SHA-1 hashes in NSEC3 records. Concerns about the load created by the computation of NSEC3 records on the DNS resolvers were already considered in the NSEC3 specifications RFC5155 and RFC9276. In February 2024, the potential of NSEC3 to exhaust DNS resolvers' resources was assigned a CVE-2023-50868, confirming that extra iterations of NSEC3 created substantial load. However, there is no published evaluation of the attack and the impact of the attack on the resolvers was not clarified.

\indent In this work we perform the first evaluation of the NSEC3-encloser attack against DNS resolver implementations and find that the NSEC3-encloser attack can still create a 72x increase in CPU instruction count, despite the victim resolver following RFC5155 recommendations in limiting hash iteration counts.
The impact of the attack varies across the different DNS resolvers, but we show that with a sufficient volume of DNS packets the attack can increase CPU load and cause packet loss. We find that at a rate of 150 malicious NSEC3 records per second, depending on the DNS implementation, the loss rate of benign DNS requests varies between 2.7\% and 30\%.
We provide a detailed description and implementation of the NSEC3-encloser attack.
We also develop the first analysis how each NSEC3 parameter impacts the load inflicted on the victim resolver during NSEC3-encloser attack.

\indent We make the code of our NSEC3-encloser attack implementation along with the zonefile and the evaluation data available for public use: \url{https://github.com/Goethe-Universitat-Cybersecurity/NSEC3-Encloser-Attack}.

\end{abstract}

%% file: 010-introduction.tex
\section{Introduction}\label{sec:introduction}
On 13 February 2024 a vulnerability,\footnote{\url{https://kb.isc.org/docs/cve-2023-50868}} termed {\em Preparing an NSEC3 closest encloser proof can exhaust CPU resources}, was registered as CVE-2023-50868 (short for Common Vulnerabilities and Exposures) in a list of publicly disclosed information security flaws. The description of the CVE says that the processing of responses sent by nameservers authoritative for DNSSEC signed zones can exploit maliciously crafted NSEC3 records to cause CPU exhaustion on a DNSSEC-validating resolver. By flooding the target resolver with queries, an adversary can trigger responses to the target resolver with specially crafted NSEC3 records exploiting this flaw. Computation of those NSEC3 records can significantly impair the resolvers' performance. In this work, we provide the first analysis of the vulnerability and an evaluation of the attack against popular DNS resolvers. We explain the impact on the resolvers' implementations using code analysis as well as monitoring of the CPU instruction count and measurements of the latency incurred on requests from benign clients. 

{\bf Vulnerabilities in proof of non-existence.} Domain Name System Security (DNSSEC) RFC4033 -- RFC4035 was designed to protect the Domain Name System (DNS) against manipulation attacks by attaching digital signatures to DNS records. The DNS resolvers can use the public keys of the corresponding domains to authenticate the DNS records that they receive in responses. To provide an authenticated proof for resources that do not exist, RFC3845 defined NSEC records, which list the hostname before and the hostname after the requested hostname. The listing of hostnames in NSEC records exposed the domains to zone enumeration attacks, discussed in RFC4470. To mitigate zone enumeration attacks, the IETF standardized NSEC version 3 (NSEC3) in RFC5155. NSEC3 computes hashes over the hostnames and the resulting NSEC3 record lists the hashed names instead of plaintext names. Nevertheless, NSEC3 too was found vulnerable to zone enumeration attacks \cite{bau2010security,goldberg2015stretching,mitchelltaking}. Although the privacy aspects of NSEC3 records were substantially explored, there was no evaluation of the performance impact of NSEC3 records on DNS resolvers. In this work, we provide the first evaluation of the performance load induced on the resolvers by attacks with specially crafted NSEC3 records, which we dub the {\em NSEC3-encloser} attack. Although the potential degradation of performance by NSEC3 records was considered in RFC5155\#\S8.3, there was no evaluation of the impact on performance by attackers and the role of the NSEC3 parameters on the effectiveness of the attack. A recently registered CVE-2023-50868 does not explain the impact of the attack on the resolvers nor provides the evaluation of the attack.

{\bf NSEC3-encloser can exhaust CPU and lead to loss.} We implement and evaluate an NSEC3-encloser attack that leads to increased CPU instruction counts on the affected resolvers, and also to loss of packets from legitimate clients. In our implementation of the attack, the NSEC3 records use the maximum number of {\tt iterations} supported by the DNS resolver implementations, which follow the recommendation counts listed in RFC5155. We experimentally observe that using salt in the calculation of hashes in NSEC3 results in a more effective attack than attacks without the salt. The reason is that the salt value creates an additional input block which leads to an increased calculation time since the blocks are processed sequentially. 
At the same time, the salt value does not substantially increase the resilience to zone enumeration attacks since, in contrast to the traditional uses of the salt in hash computations like for passwords, the hashes are implicitly salted per zone by including the domain name in the computation process. This is also stated in RFC9276, and limits the benefit of using a salt in the first place.

Our contributions can be summarized as follows:

$\bullet$ We develop a tool for automated evaluation of the CVE-2023-50868 attack, expanding on the proof-of-concept in the CVE, and providing an automated setup to generate zones and queries. Our implementation creates multiple NSEC3 configurations setting different values for NSEC3 parameters, including a novel method for maximizing the number of NSEC3 records in DNS responses and varying salt length, all of which allow for testing different aspects of the resolvers' behavior. We make our tool open-source to facilitate reproduction of our work~\cite{gruza_2024_11352869}.

$\bullet$ We provide the first evaluation of an attack that exploits NSEC3 records for creating a load on DNS resolvers. In our evaluation, we also analyze the resolvers' behavior and limits introduced in RFC5155 and explain how the resolvers react to different values of NSEC3 parameters. We find that the salt increases the load on the resolvers by 30\%, an aspect which was previously overlooked and not included in either CVE-2023-50868 or the PoC that the CVE made public. Our full fledged and automated attack evaluation allowed to identify the role of salt in increasing the CPU instruction counts on the resolvers. We also explore the limitations of the NSEC3-encloser attack, i.e., the high query rate required to load resolvers and the relatively low impact on traffic loss.

$\bullet$ We perform the first comparison of the NSEC3-encloser attack to other attacks on DNS, and explain the differences in performance and load, as well as in the vulnerabilities in resolvers' behavior that are exploited.

$\bullet$ We perform measurements of NSEC and NSEC3 configurations on DNSSEC-signed domains and find that 56\% of the domains use NSEC which is vulnerable to zone enumeration, while 41\% use NSEC3. 77\% of those NSEC3 domains use a high number of hash iterations which exposes those domains for abuse to create load on victim resolvers.

{\bf Organization.}
This paper is organized as follows.
In Section~\ref{sec:background-related-work}, we provide an overview of DNSSEC and the proof of non-existence with NSEC and NSEC3.
We provide the details of the NSEC3 attack in Section~\ref{sec:attack}.
We evaluate the NSEC3 attack in Section~\ref{sec:praceval}, demonstrating the role of the parameters in the NSEC3 record on the impact of the attack.
We measure real-world DNSSEC and NSEC/3 in Section~\ref{sc:measurementsdomains}.
Finally, we review Related Work in Section~\ref{sc:works} and conclude in Section~\ref{sec:conclusions}.

%% file: 020-background.tex
\section{Overview of DNSSEC and NSEC3}\label{sec:background-related-work}
The IETF standardized DNSSEC RFC4033 -- RFC4035 to enable DNS resolvers to detect if DNS records in responses are manipulated. The DNSSEC specification requires that the records in a zonefile are digitally signed. The zonefile contains DNS records as well as DNSSEC material, most notably DNSKEY, RRSIG, and DS records. 

DNSSEC signatures are stored in RRSIG-type DNS records. The public keys used to validate the signatures are sent in DNSKEY-type records. DS records from a parent zone are used to authenticate individual Key Signing Key (KSK) type DNSKEY records in a child zone. This is done to delegate trust from a parent zone public key to a child zone public key. DS records use the same triple (owner name, algorithm, key tag) to identify a subset of candidate DNSKEYs as RRSIGs.

In additional to cryptographically attesting the validity of DNS records, DNSSEC also enables proofs for non-existing records, enabling authenticated denial of existence. 

For this, RFC4035 defines Next Secure (NSEC) records for a precomputed denial of existence, that prove that a requested hostname does not exist. Each NSEC record contains a signed pair of consecutive hostnames, sorted canonically. Each query for a hostname not in the zonefile is answered by the nameserver with a suitable NSEC record. For instance, a query for a non-existing hostname \texttt{b.x.org} is responded with a signed NSEC record for a pair of existing hostnames sorted canonically before and after the queried hostname: \texttt{a.x.org} and \texttt{c.x.org}. The resolver can then confirm the requested hostname does not exist as the NSEC record attests no domain name exists between \texttt{a.x.org} and \texttt{c.x.org}, proving non-existence of \texttt{b.x.org}. An example of a NSEC record is given below.

{\small{
\begin{verbatim}
\\ Domain | TTL | RR type | Next hostname
   x.org    700   NSEC      a.x.org      

\\ Resource record sets
   NS SOA RRSIG NSEC DNSKEY
\end{verbatim}
}}

Research showed that NSEC was vulnerable to zone enumeration attacks \cite{bau2010security,goldberg2015stretching,mitchelltaking}. By enumerating a target zone, an adversary learns the IP addresses of all resources in the target zone. An enumerated list of resources can be exploited for other attacks, such as spam. To mitigate the threat introduced by NSEC records, RFC5155 designed NSEC3: a precomputed denial of existence.
The idea of NSEC3 is replacing clear-text hostnames with hashes, which makes zone enumeration from the names significantly harder.
The knowledge of the hashed hostname cannot be directly used for zone enumeration since cryptographic hash functions do not allow for the reconstruction of the plaintext hostname through preimage resistance. NSEC3 uses an additional record NSEC3PARAM which contains parameters for the NSEC3 validation, including the hash algorithm, the amount of iterations, and salt parameters. A single NSEC3PARAM record dictates the parameters for the entire set of NSEC3 records. This is needed to ensure that any query for a non-existent hostname maps to an NSEC3 record. The `salt' contains hexadecimal digits and is appended to the domain name to make offline dictionary attacks harder. `Iterations' indicates the number of times the hash function was computed.

The NSEC3 record contains a pair of ordered hashes. According to RFC5155, to create the NSEC3 records, the canonical hostname is hashed once and the resulting hash is re-hashed a number of times according to the iteration parameters in the NSEC3PARAM. Upon a query for a non-existent resource, the nameservers should return to the requesting resolvers a signed NSEC3 record that contains two hashes, one before the requested hostname and one after. The resolver can then hash the hostname to ensure the hashed hostname lies between the returned hashes, thereby proving the non-existence. An example of an NSEC3 record is given below.

{\small{
\begin{verbatim}
\\ Hashed domain | TTL | RR type | Algorithm
   ej23jdn4jnd...  700    NSEC3     1 (SHA1)
   
\\ Flags | Iterations | Salt       
      0       150       64ccab74...  
      
\\ Next hostname | Resource record sets
   kev723jd...     NS SOA RRSIG NSEC DNSKEY
\end{verbatim}
}}

RFC9276 defines the best current practice for setting and dealing with NSEC3 parameters, including considerations of Denial of Service (DoS) by Central Processing Unit (CPU) resource exhaustion through NSEC3 hashing. The only hash function standardized for use in NSEC3 records is SHA-1.\footnote{\url{https://www.iana.org/assignments/dnssec-nsec3-parameters/dnssec-nsec3-parameters.xhtml}}

According to RFC5155\#\S7.2, the resolvers require a proof of the closest encloser, which proves that a subdomain of the requested hostname is the closest encloser of that name. The proof consists of up to two NSEC3 records: An NSEC3 record that matches the closest (provable) encloser and an NSEC3 record that covers the ``next closer'' name to the closest encloser. The first NSEC3 record proves that the encloser exists. The second NSEC3 record proves that the possible closest encloser is the closest, and proves that the queried hostname (and any subdomains between the queried hostname and the closest encloser) does not exist. These NSEC3 RRs are collectively referred to as the ``closest encloser proof'' RFC5155. An example in RFC5155 describes the closest encloser proof for the nonexistent hostname {{\texttt{alpha.beta.gamma.example.}}}: The owner might prove that {{\texttt{gamma.example.}}} is the closest encloser. The response contains the NSEC3 record that matches {{\texttt{gamma.example.}}}, and also contains the NSEC3 record that covers {{\texttt{beta.gamma.example.}}} (which is the ``next closer'' name).

According to the specification in RFC5155 to prove the nonexistence of a hostname in a query, a closest encloser proof and an NSEC3 record covering the (nonexistent) wildcard record at the closest encloser MUST be included in the response. This collection of (up to) three NSEC3 records proves both that the queried hostname does not exist and that a wildcard that could have matched the queried hostname also does not exist; if  {{\texttt{gamma.example.}}} is the closest provable encloser to the queried hostname, then an NSEC3 record covering  {{\texttt{*.gamma.example.}}} is included in the authority section of the response.

%% file: 040-attack.tex
\section{NSEC3-Encloser Attack}\label{sec:attack}
The NSEC3-encloser attack exploits computational complexity in hash calculation for closest encloser proofs. The idea behind the attack is to set up a malicious zonefile in a valid DNSSEC signed domain, then to cause the victim DNS resolvers to issue DNS queries for a non-existent resource in the domain of the adversary. We design our attack to be fully RFC compliant; both the client requesting resolution from the victim resolver as well as the nameserver containing the malicious zonefile fully conform to all RFC requirements. The goal is to create a zonefile that maximizes both the number of hash calculations and the computation effort per single hash calculation. We construct an attack on NSEC3 instead of NSEC as the former requires hash calculations for the closest encloser proof, which significantly increases computational load compared to NSEC. 
The core aspect of the NSEC3 attack lies in the construction of the proof of non-existence with NSEC3 records, which should lead to many hash calculations in the victim resolver. The adversary requests a resource that inflicts large complexity for the resolver to prove the closest encloser.
In the following, we illustrate the attack concept with exemplary adversarial zonefiles.

\subsection{Zonefile Construction}\label{sc:zonefile}

In the configuration of the zone, we follow DNSSEC and NSEC3 standard specifications. This ensures that the zonefiles are accepted by all standard compliant resolvers.

To maximize the attack impact, the attacker needs to trigger the maximum number of hash validations in a victim resolver.
Since each NSEC3 record obtained from a DNS request results in a single hash calculation, this corresponds to maximizing the number of NSEC3 records for a given
request.
Following RFC5155, this number is limited to up to three NSEC3 records per DNS request, leading to a maximum of three hash calculations per request.
Achieving this maximum number of NSEC3 records in each resolver request requires a specific zonefile configuration, which we illustrate in Figure~\ref{fig:example_zonefile}.
For a configured zone origin, the generated zonefile consists of the following non-NSEC3 (and non-RRSIG) records:

$\triangleright$ The SOA, NS and DS records of the zone, present at the zone apex.

$\triangleright$ Two DNSKEY records, one for the KSK and one for the ZSK.

$\triangleright$ One NSEC3PARAM record at the zone apex, signaling NSEC3 usage to the authoritative nameserver.

$\triangleright$ The A record for the nameserver domain.\vspace{3pt}

The zone has two unique name entries, {\texttt{ATTACK.ER}} and {\texttt{NS1.ATTACK.ER}}.
Following specification, both of these names require an NSEC3 record, proving the existence of the Resource Record sets (RRsets) listed for the names. However, to achieve three NSEC3 records in the response for an arbitrary resolver request, this is insufficient, as any domain existence or non-existence proof would require between one and two of these NSEC3 records.
To validate an NSEC3 reply, resolvers need three different values from the nameserver: The closest encloser, proof that the ``next closer'' domain does not exist, and proof that no wildcard record exists covering the requested domain.

The closest encloser proves that a domain exists in the zone that is the nearest ancestor of the queried name. It establishes a context within which the non-existence of the queried domain can be asserted. In our example, the NSEC3 record with the hash of {{\texttt{ATTACK.ER}}} proves the existence of this hostname, and all subdomains will receive this record as their closest-encloser.

The next domain hash of an NSEC3 record provides evidence of the numerically subsequent domain name hash in the zone, confirming that no records exist between the domain name hash of an NSEC3 record and this next domain.
For example, consider a nameserver has to proof the non-existence of a domain with a hash of 0x123.
In the zone, the next smaller NSEC3 record has a hash of 0x111, with a next hash value of 0x222.
Since the requested domain hash (0x123) is larger than 0x111 but not equal to 0x222, the requested domain provably does not exist in the zone.
The nameserver must provide the NSEC3 record proving that the ``next closer'' domain (the ancestor of the queried name just below the closest encloser) does not exist.
The resolver can confirm that this domain name does not exist by validating that the next hash in the returned NSEC3 record is not the hash of the ``next closer'' domain.
By inference, the queried name cannot exist, too, since the zone does provably not include one of its ancestors.

Finally, the resolver needs to ensure that no wildcard record covers the requested domain.
The nameserver thus includes the NSEC3 record next-smaller of where the hash of the wildcard record corresponding to the level of the enclosed domain would be.
These proofs may, however, overlap. For example, if the next domain corresponds with the NSEC3 entry for the closest encloser, the nameserver will only send the overlapping entry once, reducing the resulting computational effort in the resolver, thereby weakening the attack. 

\definecolor{rtype}{RGB}{13, 13, 193}
\definecolor{rtypealt}{RGB}{194, 7, 7}
\newcommand{\rtype}[1]{{\color{rtype}{#1}}}
\newcommand{\rtypealt}[1]{{\color{rtypealt}{#1}}}

\begin{figure}[t!]
    \centering
    \sffamily 
    \begin{minipage}{0.45\textwidth}
    \setlength{\fboxsep}{5pt} 
    \setlength{\fboxrule}{1pt} 
    \fbox{
    \begin{minipage}{.94\linewidth}\raggedright
        \setlength{\parindent}{-1em}
        \setlength{\leftskip}{1.5em}
        \setlength{\parskip}{0.5em}
        \hspace{-.5em}\textbf{;; ZONE `ATTACK.ER'}
        \smallskip

        {\footnotesize\bfseries
        \indent ATTACK.ER. 0 IN \rtype{SOA} NS1.ATTACK.ER. NS1.ATTACK.ER. 0 0 0 10 0
        
        ATTACK.ER. 0 IN \rtype{NS} NS1.ATTACK.ER.

        ATTACK.ER. 0 IN \rtype{DS} 35650 7 1 e8316...

        ATTACK.ER. 0 IN \rtype{DNSKEY} 257 3 7 AwEA...\\
        \indent ATTACK.ER. 0 IN \rtype{DNSKEY} 256 3 7 AwEA...

        ATTACK.ER. 0 IN \rtype{NSEC3PARAM} 1 0 150 -

        HKHV...38AU.ATTACK.ER. 0 IN \rtypealt{NSEC3} 1 1 150 - HKHV...38B0\:\:\circled{1}

        HKHV...38B0.ATTACK.ER. 0 IN \rtypealt{NSEC3} 1 1 150 - QCQC...7U45\:\:\circled{2}

        NS1.ATTACK.ER 0 IN \rtype{A} 6.6.6.6

        QCQC...7U45.ATTACK.ER. 0 IN \rtypealt{NSEC3} 1 1 150 - SN5U...89IT A RRSIG\:\:\circled{3}

        SN5U...89IT.ATTACK.ER. 0 IN \rtypealt{NSEC3} 1 1 150 - SN5U...89IU NS SOA DS RRSIG DNSKEY NSEC3PARAM\:\:\circled{4}

        SN5U...89IU.ATTACK.ER. 0 IN \rtypealt{NSEC3} 1 1 150 - HKHV...38AU\:\:\circled{5}

        [...] ;; RRSIG records
        
        }%
    \end{minipage}%
    }
    \end{minipage}%
    \caption{Generated attack zonefile example.}
    \label{fig:example_zonefile}
\end{figure}

To force the authoritative nameserver to serve exactly three NSEC3 records to every request for a non-existing domain name and thereby maximize the impact of the attack, we develop a new scheme for NSEC3 records in the zone. The required records are described in the following.
Note that $H$ is the NSEC3 hash function used, generally SHA-1.

\renewcommand\labelenumi{(\arabic{enumi})}

\begin{enumerate}{{
    \item {$H(\texttt{ATTACK.ER})$\texttt{.ATTACK.ER} with next hash~(\ref{enum:zoneplus})\label{enum:zone}
    \item $H(\texttt{NS1.ATTACK.ER})$\texttt{.ATTACK.ER}\label{enum:ns}
    \item $(H(\texttt{ATTACK.ER}) + 1)$\texttt{.ATTACK.ER}\label{enum:zoneplus}
    \item $(H(\texttt{*.ATTACK.ER}) - 1)$\texttt{.ATTACK.ER} with next hash~(\ref{enum:wildplus})\label{enum:wildminus}
    \item $(H(\texttt{*.ATTACK.ER}) + 1)$\texttt{.ATTACK.ER}\label{enum:wildplus}}}
    
    }
\end{enumerate}

NSEC3 records (\ref{enum:zone}) and (\ref{enum:ns}) are mandatory records and thus must be included in the domain. Further, in the attack setup, the adversary will trigger resolution of a non-existent subdomain of the {\texttt{ATTACK.ER}} domain, resulting in (\ref{enum:zone}) always contained in the reply as it is the closest encloser to all requests. Note that this closest encloser NSEC3 record also includes a next-hash value. If the resolver requests a domain which is, by chance, hashed to a value directly ``after'' the {\texttt{ATTACK.ER}} domain hash, the authoritative server would detect the overlap and only send a single NSEC3 record (\ref{enum:zone}) to cover closest encloser and the next hash. To prevent this and force an additional NSEC3 record in the answer, we include an additional NSEC3 record (\ref{enum:zoneplus}) which covers the hash one larger than (\ref{enum:zone}). Thus, (\ref{enum:zone}) always has (\ref{enum:zoneplus}) as next hash and therefore never covers any other non-existent domain in the zone.
It will therefore never overlap with the required ``next closer'' domain record.

Similarly, the attacker needs to ensure that none of the above mentioned records, by chance, covers the wildcard domain name, as the resolver would then, e.g., only need to send a single record for ``next closer'' and wildcard proof.
To prevent this, a new record (\ref{enum:wildminus}) is added, with a hash value just below the hash of the wildcard domain name, as this record will now always be included to proof non-existence of the wildcard domain.
Conversely, this new record now also has a next hash value, which might by chance cover the ``next closer'' domain of the requested domain, again leading to overlap.
Therefore, a new record (\ref{enum:wildplus}) is added that ensures that the record (\ref{enum:wildminus}) only covers two hashes. Thus, for every request to a non-existent domain, the nameserver must include three NSEC3 records: \textbf{(1)} for closest encloser, then \textbf{(2), (3)} or \textbf{(5)} for the ``next closer'' proof, and finally \textbf{(4)} for wildcard proof.

\subsection{Maximizing the Impact}
Using the above described zonefile as-is only results in three hash computations. However, the impact can be increased, both by adapting the DNS request from client to resolver, and by adapting the malicious zone.

\subsubsection{Adapting the request} 
When a client requests a non-existent domain from the resolver, the resolver needs to conduct the above described checks to attest non-existence of the domain, including the check for the closest encloser. Crucially, the resolver cannot necessarily directly infer the closest encloser from the NSEC3 records.
For instance, consider a nested sub-domain {\texttt{A.B.ATTACK.ER}}. The resolver receives a hash for the closest encloser, but does not directly know if the hash is for {\texttt{A.B.ATTACK.ER}}, {\texttt{B.ATTACK.ER}}, or {\texttt{ATTACK.ER}}.
Instead, the resolver has to attempt for each candidate individually whether any of the NSEC3 records in the response proves the existence for the encloser.
The algorithm for this is listed in RFC5155.
The resolver hashes the query name and matches the resulting hash against each NSEC3 record. If none of the records fit, it has to slice away the next label and try again, repeatedly hashing and matching.
Therefore, the workload of the closest encloser proof depends on the number of labels below the closest encloser in the query name and, to a lesser degree, on the number of NSEC3 records in the nameserver response.
Maximizing these numbers can incur a significant workload of calculating hashes on the resolver.
Note that the maximum number of labels in the request is limited by the maximum request size of 255 bytes in RFC1035. 

\subsubsection{Adapting the zone} Using NSEC3 parameters in a malicious zonefile, the per-hash overhead can be greatly increased. In the following, we highlight the two NSEC3 parameters that can be manipulated to maximize impact. 

\subsubsection{Hash iteration count} NSEC3 supports hash iterations to increase computational effort for brute-forcing hash values.
Hash iterations require that the hash of a domain name is re-iterated through the respective hash function for a set number of iterations.
This mechanism, while improving security through hardening brute-force protection, can be exploited to increase computational load per calculation on the resolver, resulting in a stark increase in the number of hash calculations in the attack.
For example, if the resolver needs to calculate three hashes for the three NSEC3 records in the zone, choosing an iteration count of 100 will result in a total of 300 hash calculations.  

\subsubsection{Adding a salt} Additionally to iterations, NSEC3 also supports protection against rainbow-table attacks~\cite{kumar2013rainbow} through the addition of a salt value to the hash. The salt is added to the plaintext domain name before hashing, which prevents pre-calculation of tables of potential domain names. The salt additionally increases the computational load for hash calculations, as SHA-1 (the only currently supported hash algorithm) exhibits an increase in computation time over longer plaintext inputs. The increase in computation time stems from the underlying blocks that are used as input to the hash functions; with more blocks of plaintext, the hash function takes linearly more time. Notably, when using iterations, the salt is not only added to the first iteration of the hash function but to all subsequent inputs to the function, increasing load for each of the iterations. 

Our code-review yields that all investigated resolvers support both the hash iterations and the salting, following RFC specification. An exemplary implementation of the hash function in Unbound DNS resolver is given in code Listing \ref{lst:nsec3-iterations-unbound}.

\newpage

\begin{lstlisting}[language=c++, caption=Source code for NSEC3 iterations in Unbound., label={lst:nsec3-iterations-unbound}]
nsec3_calc_hash(struct regional* region,
      sldns_buffer* buf, struct nsec3_cached_hash* c) {
  // [...] Init buffers and do sanity checks

  // Write dname and salt to buffer
  sldns_buffer_write(buf, c->dname, c->dname_len);
  sldns_buffer_write(buf, salt, saltlen);

  // Calculate first hash from buffer content
  (void)secalgo_nsec3_hash(algo,
      (unsigned char*)sldns_buffer_begin(buf),
      sldns_buffer_limit(buf),
      (unsigned char*)c->hash);

  for(i=0; i<iter; i++) { // Iterate through number
    // Insert previous hash and salt into buffer
    sldns_buffer_clear(buf);
    sldns_buffer_write(buf, c->hash, c->hash_len);
    sldns_buffer_write(buf, salt, saltlen);

    // Calculate hash from buffer content
    (void)secalgo_nsec3_hash(algo,
        (unsigned char*)sldns_buffer_begin(buf),
        sldns_buffer_limit(buf),
        (unsigned char*)c->hash);
  }
}
\end{lstlisting}

The code snippet shows how the NSEC3 iterations are performed.
The hash is calculated and written into the result.
Then, a for-loop is entered which continuously writes the result of the previous hash calculation into a clear buffer, adds the salt and calculates the hash again, as long as the iteration count is below the limit.
The code-example shows that Unbound, like all investigated resolvers, conforms to the specification in iterating the hash and adding the salt in each iteration.

\begin{table}[t!]
\renewcommand{\arraystretch}{1.0}
\centering
\footnotesize
    \begin{tabular}{l||c|c|c}
        \hline
        \textbf{Resolver} & \multicolumn{3}{|c|}{\textbf{Iteration Limits}} \\\cline{2-4}
     & Patched version & 50 & 150\\ 
        \hline
        Unbound & 1.19.1 &--- & 1.13.2 \\
        Bind9 & 9.18.24, 9.19.19 & 9.18.24, 9.19.19 & 9.17.13 \\
        &9.16.48 & &9.16.16 \\
        PowerDNS & 5.0.0 & 5.0.0 & 4.5.2 \\
        Knot Resolver & 5.7.1 & 5.7.1 & 5.3.1 \\
        \hline
    \end{tabular}
    
    \caption{The limits introduced across resolvers over time.}
    \label{tab:resolver_nsec3_iteration_limits}
\end{table}

\textbf{Practical limits to iterations.} The standard provides recommendations to the number of iterations a resolver may allow on a given NSEC3 record. 
We find from code review that these values are observed only in some resolvers; a subset of resolvers do not enforce these limits, while other resolvers set stricter limits in their standard configuration. This is not surprising, as RFC9276 encourages resolvers to choose their own limits to a value they seem adequate for current deployments. A detailed overview of enforced iteration limits in different resolver versions is presented in Table \ref{tab:resolver_nsec3_iteration_limits}. 

\textbf{Practical limits to salt length.} Generally, a longer salt value allows for longer calculation time of a given hash. However, the maximum length of the salt is limited by the available space of the salt field in the NSEC3PARAM record, only allowing up to 255 bytes of data for the salt. We find from code review that all resolvers allow this maximum salt length, with no resolver enforcing stricter length limits. 

{\em Thus the maximum attack impact can be achieved by querying the resolver with a deeply nested sub-domain, configure the nameserver to always deliver all three NSEC3 records, and using both the maximum number of iterations allowed by the resolver, and the longest possible salt length of 255 byte.}

\subsubsection{Generating the zonefile}
To test different zone configurations with differing values for the NSEC3 parameters, we develop a script that automatically generates zonefiles from a singular JSON configuration file. We make the script publicly available to facilitate reproduction of our work~\cite{gruza_2024_11352869}.
This configuration file used in the script specifies the individual zones, the cryptographic parameters, such as key size and NSEC3 iterations, nameservers, TTL values, and relationship between the zones.
The generation script written in Python parses a configuration, generates the defined records, creates all relevant DNSSEC signature and key records, and exports each zone to a file to be hosted by a nameserver implementation.

%% file: 050-impact.tex
\section{Evaluation of the Attack}\label{sec:praceval}

To practically evaluate the impact of the attack, we deploy the resolvers and a nameserver with the attack zones in a local isolated setup.
We send attack queries to the resolvers and measure the impact of the attack under different scenarios.
Section~\ref{sc:setup} describes the test setup, Section~\ref{sc:parameter_evaluation} illustrates the influence of different parameters on the impact of the attack, and Section~\ref{sc:resolver_comparison} delves into comparing the impact of the attack between different resolvers, highlighting differences in implementations that cause different reactions to attack requests.
Finally, in Section~\ref{sc:effect_on_clients}, we show that the attack can sufficiently stall resolvers to cause a drop of benign client queries.

\subsection{Setup}\label{sc:setup}
\begin{table}[tb]
\renewcommand{\arraystretch}{1.0}
\centering
\footnotesize
    \begin{tabular}{c|c|r}
        \hline
        Resolver & Version & Iteration Limit \\
        \hline
        Bind9 & 9.16.1 & RFC5155 \\
        Bind9 & 9.18.12 & 150 \\
        Unbound & 1.17.1 & 150 \\
        PowerDNS & 4.8.2 & 150 \\
        Knot & 5.6.0 & 150 \\
        \hline
    \end{tabular}
    \caption{Resolver versions and iterations limits in the test setup.}
    \label{tab:setup_resolvers}
\end{table}

We deploy the five resolvers in Table~\ref{tab:setup_resolvers} as Docker containers communicating via a network bridge with our nameserver for the attack requests, and the internet for benign requests.
We additionally include the older Bind9 version 9.16.1 in our test environment to compare the impact of the (historic) iteration count limits defined in RFC5155 to the lower limits adopted by the current implementations.
To serve the attacker zones, we set up an NSD 4.6.1 authoritative nameserver on our local network which serves the generated zonefiles.
This ensures that we accurately measure the attack impact on the resolvers, since the per-query overhead introduced by the authoritative nameserver is negligible. The nameserver is not reachable from the internet. 

We generate and include zonefiles for different combinations of parameters in NSD for each test, each having a unique identifier as part of the domain name.
The zones are generated as child zones {\texttt{EXii.NSEC3.EXAMPLE.ORG}} to a parent zone {{\texttt{NSEC3.EXAMPLE.ORG}}}, where {{\texttt{ii}}} is the two-digit zone identifier.
It is unrealistic that an attacker can control zones at the domain tree root or some top-level domain, but since the impact of the attack depends on the length of the zone domain name, we select the reasonable-length domain name {{\texttt{NSEC3.EXAMPLE.ORG}}}.
The parent zone contains signed DS records with the digest of the child zone KSK's, i.e., the zone has a complete and valid DNSSEC configuration and follows RFC specification.

Since the wire-format of the child zone domain is 24 bytes (including the root label), there remain 231 bytes for additional labels in an attacker query QNAME.
We use a randomly chosen 4-byte label as the non-existent subdomain for the attack to prevent the resolver from answering queries from the caches.
This effectively leaves 226 bytes for additional labels.
Hence, the attack query names to the resolvers have the following format, resulting in 115 sub-labels:
\begin{equation*}
    (\texttt{A.})^{113}\texttt{.abcd.EXii.NSEC3.EXAMPLE.ORG}
\end{equation*}

Each resolver is configured to query the local NSD authoritative nameserver for any queries to \texttt{NSEC3.EXAMPLE.ORG} with the zone's keys added to the set of trusted keys of the resolvers.
Furthermore, the resolvers have DNSSEC validation enabled and are run single threaded.

Our test setup is running Ubuntu 22.04 with a 12th Gen Intel\textsuperscript{\textregistered} Core{\texttrademark} i7-1280P CPU at 4.8GHz.

\subsection{Comparison of Attack Parameters}\label{sc:parameter_evaluation}

To compare the impact of the attack parameters, we execute the resolvers in a controlled environment and measure the attacker-induced CPU load for different rates of attacker queries per second and different parameter configurations.
In our analysis, we identify how specific values for configurable parameters influence the CPU exhaustion impact on the resolvers, illustrating how to maximize attack impact as well as giving a numerical basis to choose appropriate limits for attack mitigations. 

Our analysis includes key sizes, the number of NSEC3 iterations, and the length of the NSEC3 hash, influenced over the salt length. 
Each test case includes an incremental increase of the rate of attacker requests on the resolver to illustrate resolver behavior both under small scale and heavy attack. 

We conduct multiple tests to find the ideal rate for increasing the attack rate and the maximum rate of attack in the experiments. We find increasing the attack rate too quickly does not allow to distinguish the impact of a specific rate from natural fluctuations in CPU load resulting from CPU scheduling, while increasing it too slow wastes measurement time. Following our evaluation, we find increasing the attack rate every 3s as a suitable compromise. To identify a suitable maximum attack rate for the experiment, we continuously increase the rate of attack until we see artifacts caused from the experiment hardware struggling to keep up with sending enough requests to the nameserver. We find a value of 150 requests per second as a suitable maximum value were we did not observe any kernel- or hardware-induced artifacts in our measurement. A value of 150 requests per second is sufficient to cause 100\% CPU load in all investigated resolvers.
Finally, we choose to increase the attack rate with a delta of 10/3s to cause a observable difference between measurement steps, while also keeping the measurement fine-grained enough to see detailed effects at different steps. 

For Bind9.16.1, which poses no strict NSEC3 iteration limit and therefore enables a much higher attack impact per request, we reduce the attack rate to enable similar fine-grained insights. 
We identify an attack rate delta of $0.5$/s and an upper bound of $7.5$/s suitable for our setup.

In our experiments, we find that the impact of different parameters is  similar between the investigated resolvers. We will thus in the following section focus on the parameter impact on Unbound 1.17.1 and Bind9.16.1. The differences between resolvers will be discussed in Section~\ref{sc:resolver_comparison}.

\subsubsection{Key Size}

\begin{figure}[tb]
    \centering
    \begin{subfigure}{.45\textwidth}
        \includegraphics[trim=6mm 6mm 6mm 6mm, width=\textwidth, clip]{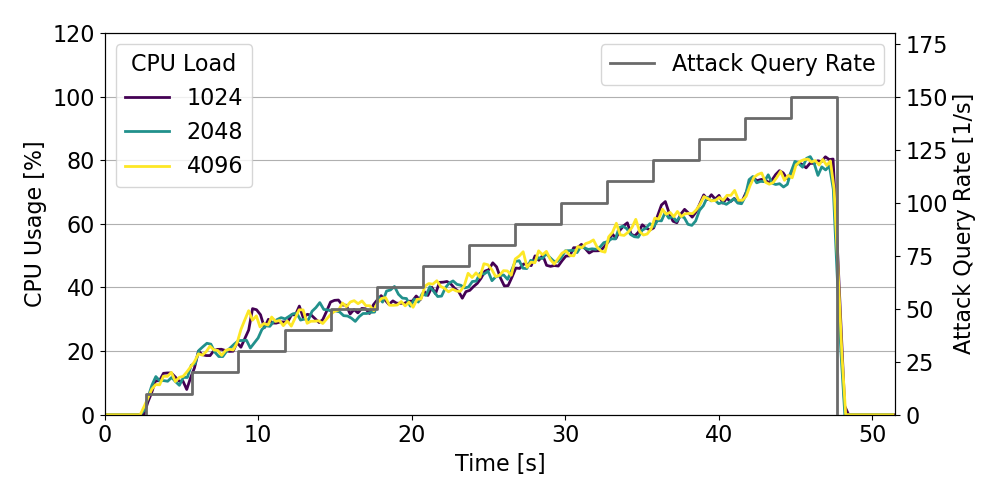}
        \caption{Unbound}
        \label{fig:parameter_keysize_unbound}
    \end{subfigure}
    \hfill
    \begin{subfigure}{.45\textwidth}
        \includegraphics[trim=6mm 6mm 6mm 6mm, width=\textwidth, clip]{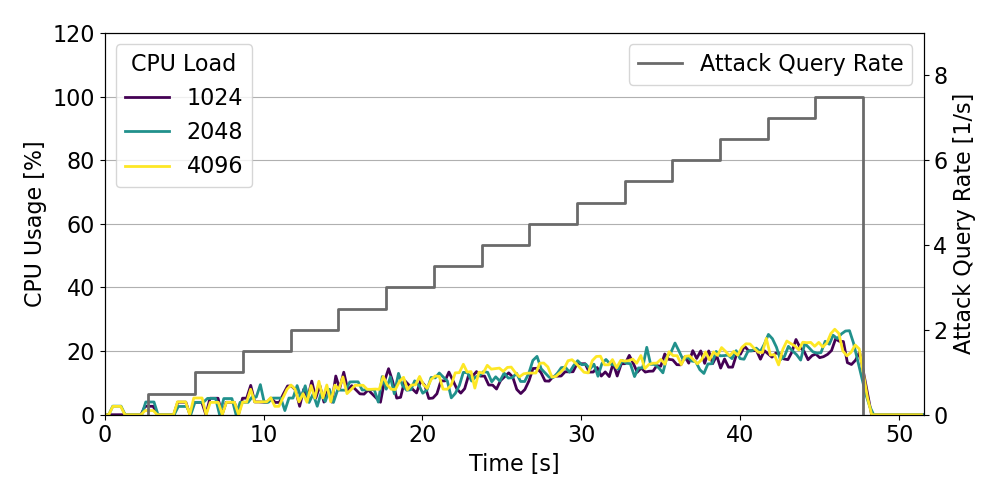}
        \caption{Bind9.16.1}
        \label{fig:parameter_keysize_bind9_16_1_low}
    \end{subfigure}
    \caption{Comparison of CPU workload for different key sizes.}
    \label{fig:parameter_keysize}
\end{figure}

While no NSEC3 parameter per se, the key size influences the maximum allowed number of NSEC3 iterations as defined in RFC5155.
We do not expect that the key size has a significant impact on the induced CPU work load since the load stems from the high number of hash calculations and not signature validation.
Nevertheless, we evaluate whether this assumption holds for the tested resolvers.
For this test, we fix the NSEC3 iterations at 150 and the salt length to 0 and compare the three different supported RSA key sizes of 1024, 2048, and 4096. 
The results are plotted in Figure~\ref{fig:parameter_keysize}.
As expected, there is no significant deviation between the three curves in CPU load.
Hence, for the subsequent tests, we use the key size 4096 as it allows for a much larger range of NSEC3 iteration values.

\subsubsection{NSEC3 Iterations}

\begin{figure}[tb]
    \centering
    \begin{subfigure}{.47\textwidth}
        \includegraphics[trim=6mm 6mm 6mm 6mm, width=\textwidth, clip]{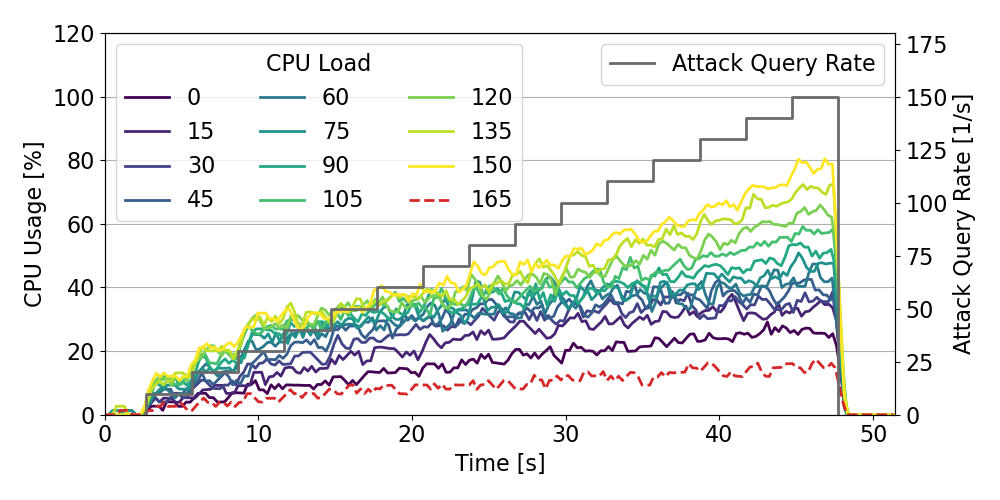}
        \caption{Unbound}
        \label{fig:parameter_iterations_unbound}
    \end{subfigure}
    \hfill
    \begin{subfigure}{.47\textwidth}
        \includegraphics[trim=6mm 6mm 6mm 6mm, width=\textwidth, clip]{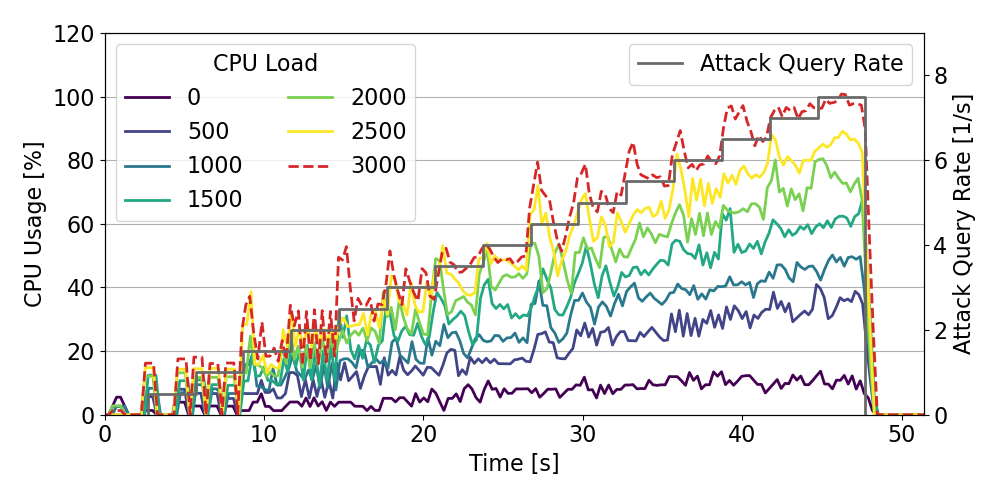}
        \caption{Bind9.16.1}
        \label{fig:parameter_iterations_bind9_16_1_low}
    \end{subfigure}
    \caption{CPU workload for different NSEC3 iteration counts.}
    \label{fig:parameter_iterations}
\end{figure}

For Unbound, we evaluate different NSEC3 iteration counts ranging between 0 and 165 in Figure~\ref{fig:parameter_iterations_unbound}.
We observe a clear correlation between higher iteration counts and larger induced workload which approaches a linear distribution as the attack query rate increases.
The exception is 165 iterations which shows loads well below all other measurements.
This is because the evaluated version of Unbound has a pre-configured limit of 150 NSEC3 iterations and disregards the zone with a higher iteration count as bogus without further validating the NSEC3 records it receives from the authoritative nameserver.
Since processing the queries and validating the signatures has some constant overhead, an iteration count of 0 incurs more overhead than the rate 165.

In the case of Bind9.16.1, no limits are enforced for the iteration values.
As evident in Figure~\ref{fig:parameter_iterations_bind9_16_1_low}, this allows us to query zones with iteration counts well above the 150 limit of all other tested resolvers.
More significantly, we can use values above the $2500$ iteration limit of RFC5155 which illustrates a significant vulnerability in this version of the resolver.
At attack rates as low as 7.5 queries per second, we are able to max out the CPU load at 100\% for this iteration limit.
But, even for the standardized $2500$ iterations, there is a significant load on the resolver, reaching up to 90\% at an attack rate of $7.5$/s. Thus, higher iteration counts can significantly increase the impact of the attack on resolvers.

\subsubsection{NSEC3 Salt Length}

\begin{figure}[tb]
    \centering
    \begin{subfigure}{.45\textwidth}
        \includegraphics[trim=6mm 6mm 6mm 6mm, width=\textwidth, clip]{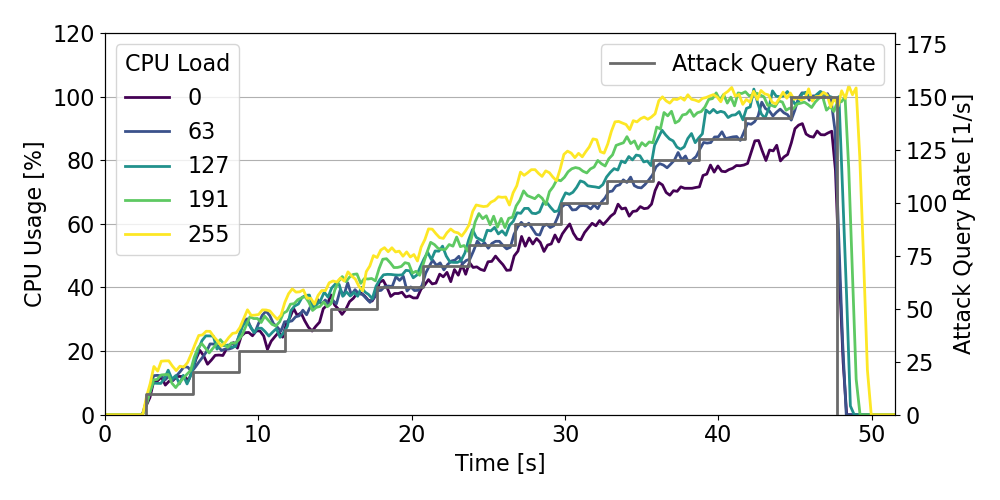}
        \caption{Unbound}
        \label{fig:parameter_salt_unbound}
    \end{subfigure}
    \hfill
    \begin{subfigure}{.45\textwidth}
        \includegraphics[trim=6mm 6mm 6mm 6mm, width=\textwidth, clip]{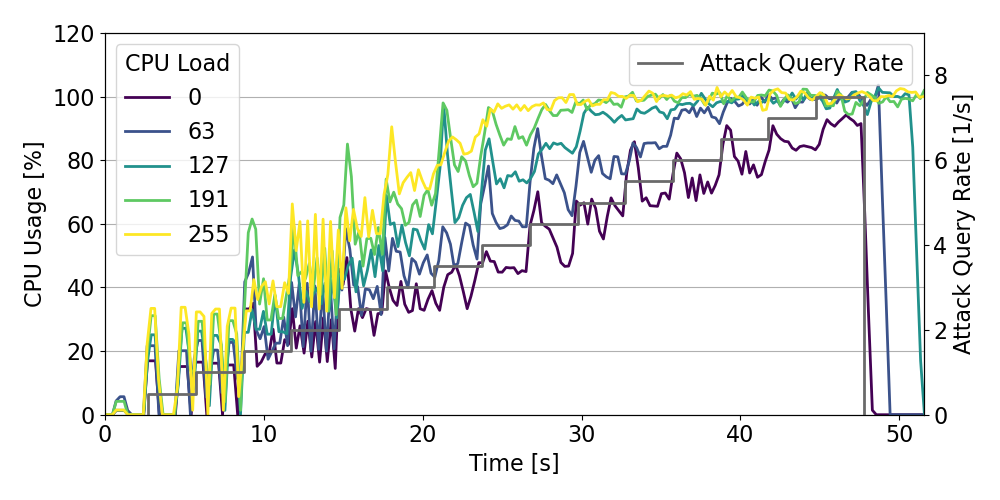}
        \caption{Bind9.16.1}
        \label{fig:parameter_salt_bind9_16_1_low}
    \end{subfigure}
    \caption{CPU workload for different NSEC3 salt lengths.}
    \label{fig:parameter_salt}
\end{figure}

Next, we compare different salt lengths in Figure~\ref{fig:parameter_salt}.
In this test, we use a key size of 4096 and set the NSEC3 iterations to the most impactful RFC5155-conform value of 150 for Unbound and $2500$ for Bind9.16.1, respectively.
For Unbound, we measure an increase of CPU load by approximately one third and for Bind9.16.1 by about one half when increasing the length of the salt from 0 to 255, with the load of the intermediate values distributed uniformly in-between.
This is to be expected from the way the NSEC3 hashes are calculated.
Since the salt is appended to the hashed domain/digest at each iteration, the additional workload of longer inputs to the hash function applies to every iteration of the hash function.
SHA-1 is a Merkle-Damg{\aa}rd hash function, hence, the calculation overhead grows roughly linearly with the number of blocks the hash function is calculated on.
With a block size of 512 bit, every 64 bytes added to the hash function input require one more calculation of the SHA-1 hash function to compute the digest.
Thus, a longer salt multiplies the total load on the resolver for each NSEC3 hash calculation by the number of blocks added through the concatenation of the salt to the digest per hash function execution.
Overall, the increased load causes the CPU load to max out at 100\% for Unbound at an attack rate of 110/s and Bind9.16.1 at $4.5$/s for a salt length of 255 bytes.

\subsection{Comparison of Resolvers}\label{sc:resolver_comparison}

\begin{figure}[tb]
    \centering
    \begin{subfigure}{.47\textwidth}
        \includegraphics[trim=6mm 6mm 6mm 6mm, width=\textwidth, clip]{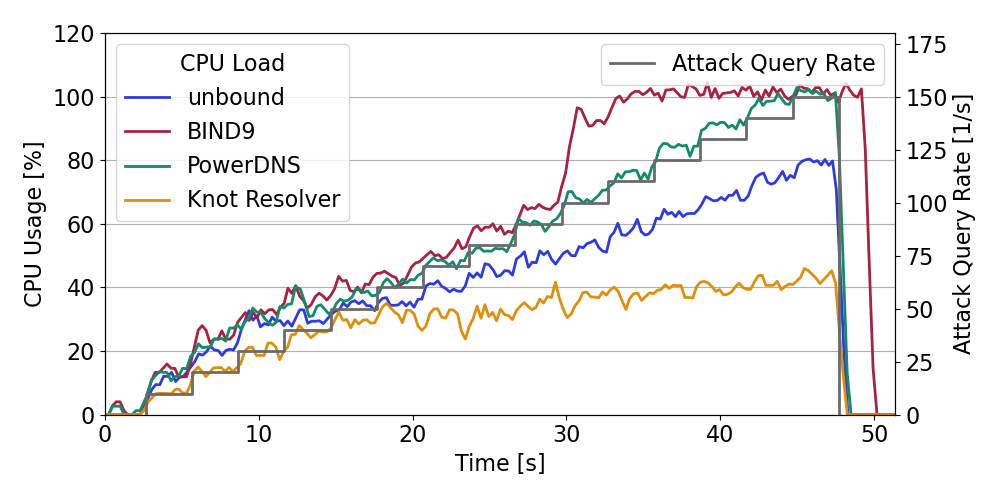}
        \caption{150 iterations, 0 byte salt}
        \label{fig:cpu_rps_resolver_comparison_0}
    \end{subfigure}
    \hfill
    \begin{subfigure}{.47\textwidth}
        \includegraphics[trim=6mm 6mm 6mm 6mm, width=\textwidth, clip]{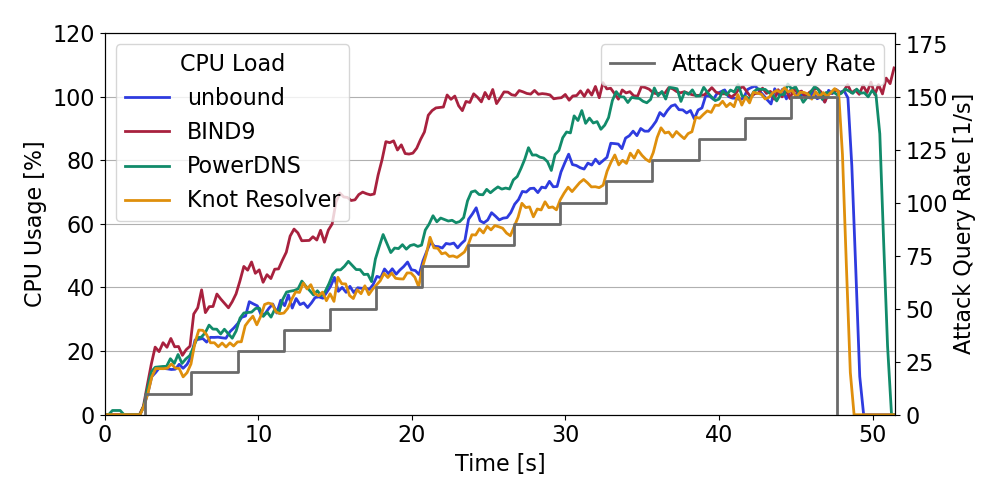}
        \caption{150 iterations, 255 byte salt}
        \label{fig:cpu_rps_resolver_comparison_255}
    \end{subfigure}
    \caption{Comparison of CPU workload between resolvers}
    \label{fig:cpu_rps_resolver_comparison}
\end{figure}

In this section, we compare the CPU load of the resolvers under the most effective parameter choices.
Since the high iteration limit in Bind9.16.1 represents a special case, we limit the comparative analysis to the resolvers with an iteration limit of 150: Bind9.18.12, Unbound 1.17.1, PowerDNS 4.8.2, and Knot 5.6.0.
As in the previous section, we execute an attack with an incrementally increasing attack rate of up to 150/s and measure the induced CPU load.
We fix the zone NSEC3 iterations at 150 and repeat the test with salt lengths 0 and 255.
The test results are illustrated in Figure~\ref{fig:cpu_rps_resolver_comparison}.

\subsubsection{Salt Length~0}
Figure~\ref{fig:cpu_rps_resolver_comparison_0} plots the CPU rates of all resolvers with a salt length of 0.
We can observe a clear differentiation of loads between the resolvers with only Bind9.18.12 and PowerDNS reaching the 100\% CPU limit before the attack rate is maxed out, at 110 and 150 packets per second, respectively.
Furthermore, we observe that Bind9.18.12 remains at 100\% CPU activity for 2 more seconds after the attack has concluded, indicating that the resolver is falling behind processing the queries in real time.
Notably, Knot is able to process the attack queries more effectively, only reaching a workload of up to 50\% during the test.

\subsubsection{Salt Length~255}
For the test case with the 255 byte salt, we illustrate the measured CPU load in Figure~\ref{fig:cpu_rps_resolver_comparison_255}.
In this scenario, all resolvers max out at 100\% CPU load before the limit of 150 attack queries per second is reached.
Bind9.18.12 reaches full load at 80/s, PowerDNS at 110/s, Unbound at 130/s, and Knot at 140/s.
This confirms that the NSEC3 salt has a significant effect on the impact of the attack on all resolvers, roughly increasing the load by a third and --- in the case of Knot -- up to one half.
Once more, we observe continuing CPU load after the attack has concluded, this time for all resolvers.
The time of continued stalling correlates with how early in the attack the full CPU load is reached because, once rates continue to rise above the rate at which the CPU is at 100\%, the resolver is unable to process the queries at the same rate as there are new incoming queries.
Bind9.18.12 continues processing queries until after the measurement has concluded. 

We can thus confirm that all examined resolvers are vulnerable to the attack.
Knot generally performs best when stressed under the resource exhaustion attack for both attack configurations, while Bind9.18.12 shows the greatest vulnerability to the attack in terms of CPU load.
In general, the effectiveness of the attack scales linearly with the attack query rate per second.

\subsection{Effect on Benign Clients}\label{sc:effect_on_clients}

\begin{figure*}[tb]
    \centering
    \begin{subfigure}{.48\textwidth}
        \includegraphics[trim=6mm 6mm 6mm 6mm, width=\textwidth, clip]{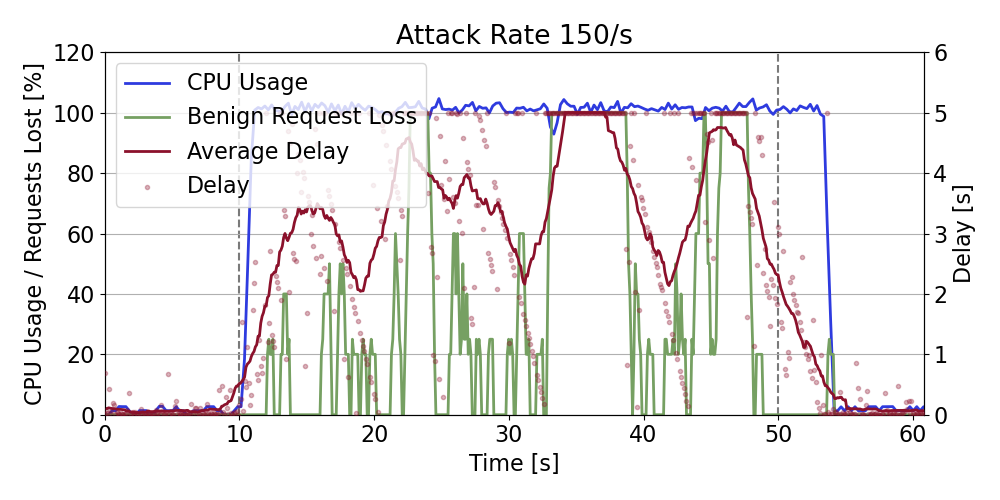}
        \caption{Unbound attacked with rate 150/s}
        \label{fig:attack_benign_150_unbound}
    \end{subfigure}
    \hfill
    \begin{subfigure}{.48\textwidth}
        \includegraphics[trim=6mm 6mm 6mm 6mm, width=\textwidth, clip]{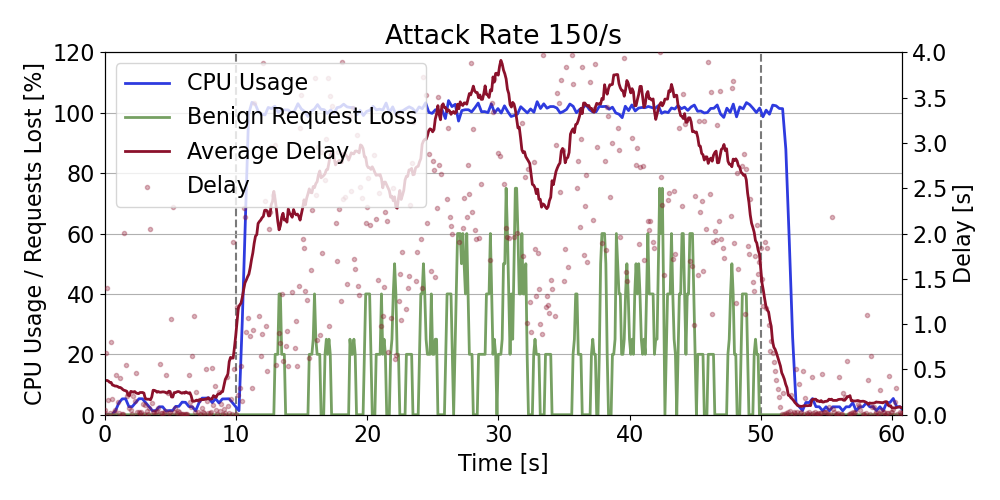}
        \caption{Knot attacked with rate 150/s}
        \label{fig:attack_benign_150_knot}
    \end{subfigure}
    \hfill
    \begin{subfigure}{.48\textwidth}
        \includegraphics[trim=6mm 6mm 6mm 6mm, width=\textwidth, clip]{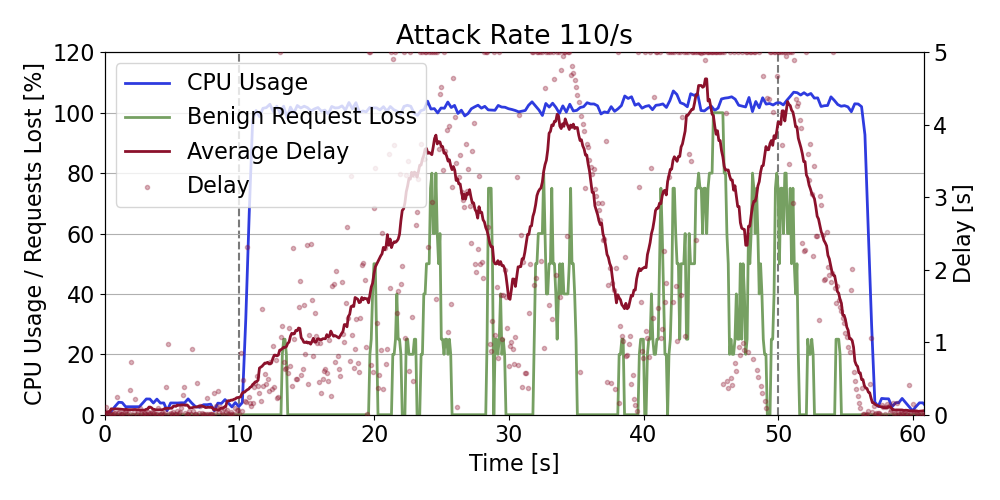}
        \caption{Bind9.18.12 attacked with rate 110/s}
        \label{fig:attack_benign_110_bind9}
    \end{subfigure}
    \hfill
    \begin{subfigure}{.48\textwidth}
        \includegraphics[trim=6mm 6mm 6mm 6mm, width=\textwidth, clip]{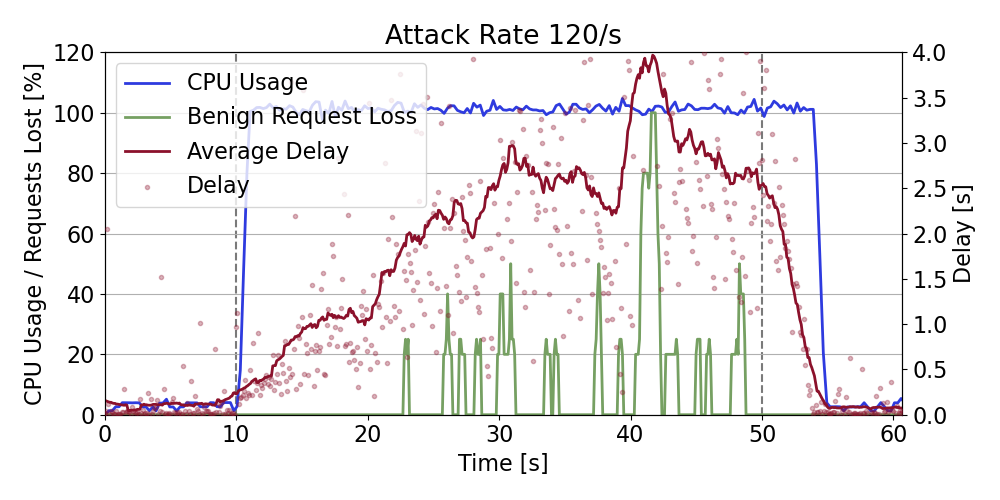}
        \caption{PowerDNS attacked with rate 120/s}
        \label{fig:attack_benign_120_powerdns}
    \end{subfigure}
    \caption{Comparison benign query delays and drops with 150 iterations and 255 byte salt}
    \label{fig:attack_benign}
\end{figure*}

\begin{table}[t!]
    \renewcommand{\arraystretch}{1.0}
\centering
\footnotesize
    \begin{tabular}{c|rrr}
        \hline
        Resolver & Attack Rate & Total Loss Rate & Adjusted Loss Rate$^*$ \\
        \hline
        Bind9.18.12 & 150/s &  $5.10\%$ &  $7.01\%$ \\
        Bind9.18.12 & 110/s & $16.42\%$ & $22.99\%$ \\
        Unbound     & 150/s & $24.75\%$ & $34.66\%$ \\
        PowerDNS    & 150/s &  $1.97\%$ &  $2.76\%$ \\
        PowerDNS    & 120/s &  $5.62\%$ &  $7.87\%$ \\
        Knot        & 150/s & $12.87\%$ & $18.01\%$ \\
        \hline
    \end{tabular}\\
    (${}^*$Total loss rate relative to the attack duration)
    \caption{Measured client request loss rate with an attack rate of 150/s over 40s, 150 iterations, and 255 byte salt.}
    \vspace{-10pt}
    \label{tab:client_loss_rate_150}
\end{table}

Having established that high query rates are required for achieving high CPU load on the resolvers, the question remains whether the attack can be used to sufficiently stall the resolvers such that they fail to answer benign client queries.
We evaluate this by continuously sending client queries at a rate of 10/s to the resolvers while simultaneously attacking the resolver with the NSEC3-encloser attack.
The clients query unique uncached records from the resolvers and log whether they receive a reply.
After 5s, we consider a client request timed out, i.e., too old to be of value to the client and therefore lost.
This is in line with the timeouts used by dig\footnote{\url{https://linux.die.net/man/1/dig}} and glibc.\footnote{\url{https://linux.die.net/man/5/resolv.conf}}
Figure~\ref{fig:attack_benign} shows the results for all tested resolvers, Table~\ref{tab:client_loss_rate_150} lists the measured client loss rates.

We measure the resolvers at attack rates of up to 150/s, starting the attack 10s into the test and executing it for 40s.
For both Unbound (Figure~\ref{fig:attack_benign_150_unbound}) and Knot (Figure~\ref{fig:attack_benign_150_knot}), we achieve adjusted loss rates --- the total loss rate during the entire test relative to the attack time --- of $34.66\%$ and $18.01\%$, respectively.
For Bind9.18.12, 150/s is well above the attack rate at which CPU utilization reaches 100\%, hence, the high number of stalled NSEC3 validations tend rapidly exhaust kernel and hardware resources and interfere with the measurement results yielding an adjusted loss rate of $7.01\%$.
Bind9.18.12 reaches a peak adjusted loss rate of $22.99\%$ at the rate of 110/s (Figure~\ref{fig:attack_benign_110_bind9}).
Similarly, PowerDNS, when attacked at 150/s, reaches a point where there are too many stalled attacker queries leading to lower loss rates in our setup.
The evaluated peak rate for PowerDNS is 120/s where we measure a loss of up to $7.87\%$ of queries at 100\% CPU utilization (Figure~\ref{fig:attack_benign_120_powerdns}).

The results show that, even with full CPU exhaustion, the attack achieves no full client query loss, i.e., no comprehensive DoS.
The key limitation of the attack is that every individual attacker query only causes a relatively minor load on the resolver, leaving ample opportunities to process and reply to client queries in-between the attacker-induced stalling periods.

\hfill

We measure the number of instructions of all resolvers for 2000 queries over a span of 40s for uncached benign queries and attack queries.
In their blog post,\footnote{https://www.isc.org/blogs/2024-bind-security-release} the developers from the Internet Systems Consortium (ISC) mention that the discovery of the NSEC3-encloser attack enables scaling the attack to 125 times as previously thought possible.
In theory, the overhead of one attack query is made up of some constant portion (e.g., for querying the authoritative nameserver and verifying the signatures) and the hash calculation.
The latter is dependent on the number of iterations, multiplied by the number of enclosed labels in the request (up to 125), roughly multiplied again by the hash operations incurred by hashing the digest plus salt (up to 4 additional hash blocks leading to a factor of approximately 3--5).
This leads to an increase of instructions by a factor of up to $125 \cdot 5 = 625$ compared to a single query with a high NSEC3 iteration count and no additional labels/salt.
In practice, hash operations are relatively cheap in terms of instructions, especially compared to asymmetric cryptography.
Hence, compared to an uncached benign query, which incurs considerable overhead through recursive querying of nameservers, retrieving keys, and validating signatures, we measure an increase of instructions by a factor of 72 for Unbound, 41 for Bind9.18.12, 33 for PowerDNS, and 13 for Knot.
The high factor for Unbound is mostly due to the low number of instructions for the benign queries, which is on average 65\% lower compared to the other resolvers.

\subsection{Comparison to PoC in CVE-2023-50868}
Following our evaluation, we also look into CVE-2023-50868, which made the NSEC3 vulnerability public and contains a Proof of Concept (PoC) implementation of the attack.

To the best of our knowledge, neither the CVE-2023-50868, nor related blog posts contain any detailed evaluations of the impact of the attack on different resolvers. We contribute this evaluation, showing that resolvers differ in their vulnerability to the attack. For example, we find that Unbound is more vulnerable to the attack due to its internal scheduling of NSEC3 compared to e.g., PowerDNS.

We further identify the impact of different NSEC3 parameters on the severity of the attack. The PoC correctly identifies that maximizing the iteration count greatly improves impact on resolvers, which we confirm in our evaluations. However, the PoC lacks utilization of a salt value, which we show to also substantially increase the attack impact. Since salts extend the length of the hash-function input, they increase the required computation in every iteration of the hash, significantly increasing effort for the resolver. 

We experimentally demonstrate that a query rate in the low hundreds is sufficient to exhaust a single CPU core on unmitigated, open-source resolver implementations at varying degrees. Using the attack, we were not able to achieve full DoS on any resolver.
Our findings illustrate that the attack is not as powerful in stalling resolvers as other attacks, such as KeyTrap \cite{keytrap} and find that this is mainly due to the linear scaling of the workload induced relative to the attacker queries, compared to a quadratic increase in load for KeyTrap. However, an attacker can still use the NSEC3 to inflict harm on resolvers and achieve a degradation of service for benign clients using the victim resolver.

%% file: 060-meaurements.tex
\section{Measurements of Signed Domains}\label{sc:measurementsdomains}

\begin{figure}[tb]
    \centering
        \includegraphics[trim=0mm 0mm 0mm 3mm, width=.45\textwidth, clip]{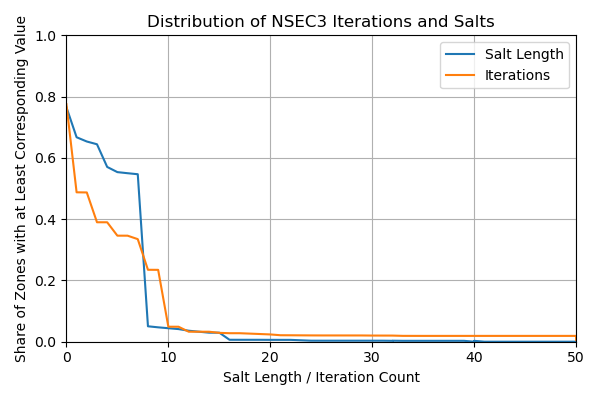}
        \caption{Share of zones which meet or exceed the configured Salt Length / Iteration Count in signed DNS zones.}
        \label{fig:distrzones}
\end{figure}

RFC9276 raises the best practice of omitting the use of both hash iterations and salts.
We measured how NSEC3 is used in domains on the Internet and investigate their NSEC3 parameter configurations.
To shed a light on how domains conform to RFC9276 and whether they use NSEC3 parameters which are suitable to be exploited in an attack,
we next quantify how many domains on the Internet use NSEC3 and which parameter configurations they employ.
During the week following 2024-03-10, we queried the nameservers of the Tranco Top-1M domains\footnote{\url{https://tranco-list.eu/list/Z333G/1000000}} for the SOA, DNSKEY as well as DS records (located at the parent) and analyzed the DNSSEC configurations of the domains they serve.
To collect information on the NSEC version and parameters used by the domains, as they are presented to the resolvers, we additionally issued queries for the records PTR-type RFC2317 at the according Tranco domain names.
PTR-type records are used for reverse-mapping IP addresses to domain names and are most commonly located below the {\texttt{IN-ADDR.ARPA.}} domain.
Therefore, we expect negative responses for these queries, indicating that no such resource exists.
Our evaluations confirm that this methodology yields negative responses, i.e., containing either first-version NSEC or NSEC3 records, for 98.15\% of the signed domains.
We find $66\,339$ ($6.63\%$) of the Tranco Top1M domains to be signed.
Out of these, $27\,761$ ($41.85\%$) use NSEC3 while $37\,354$ ($56.31\%$) use NSEC in its first version.
$21\,522$ ($77.53\%$) of the domains using NSEC3 send records with an iterations count field value higher than $0$, with a median of $5$ iterations and a maximum of $500$ iterations, while $21\,248$ ($76.54\%$) of the domains utilizing NSEC3 employ a salt.
Where employed, the median salt length is $8$ bytes and the maximum we find in our dataset is $64$ bytes.
We show the share of zones with salt lengths and iteration counts greater or equal to the respective value on the x-axis in Figure~\ref{fig:distrzones}. 
The combination of both parameters, which imposes the highest NSEC3 hashing burden on resolvers is $500$ iterations with a salt of $16$ bytes length.
According to the results of our evaluations, these domains can impose substantial load on the resolvers even with benign responses. Such domains could potentially be abused by adversaries to degrade the service of a vulnerable resolver by employing a moderate volume of malicious queries per second.

%% file: 061-works.tex
\section{Related Work}\label{sc:works}
DNS has a long history of Denial of Service (DoS) attacks which exploit different aspects of the DNS protocol to launch attacks against the DNS servers \cite{bushart2018dns,DBLP:conf/imc/MouraCHWH20,afek2020nxnsattack,afek2023nrdelegationattack}. Many of the attacks exploit a lack of limits on the functionalities performed by the DNS servers. For instance, \cite{bushart2018dns} create a chain of CNAME records and force DNS resolvers to perform deep name resolutions, hence overloading the target victim authoritative nameserver with requests and achieving an amplification of 8.51. NXNSAttack \cite{afek2020nxnsattack} exploited a vulnerability that generated a flood of queries between the recursive resolver and the authoritative server creating a load on them both. Subsequently \cite{afek2023nrdelegationattack} showed how to exploit delegations to an unresponsive authoritative server to cause computational load on DNS resolvers. Their attack differs from the NSEC3 attack in that they use plain DNS instead of DNSSEC, and create computational load through memory lookups and IO overhead instead of computational effort. Their attack achieves a higher instruction count amplification of 5600x over 70x with NSEC3. Still, the NRDelegation attack requires a high attack traffic volume of 500 requests per second to achieve substantial degradation of service, likely because the attack includes many IO operations, which allow the resolver to answer benign queries while waiting for IO responses. This explains why the NSEC3 attack, while resulting in a smaller instruction count amplification, can still achieve comparable degradation of service to NRDelegation. 

The concept of complexity attacks on DNSSEC, specifically exploiting signature validations and hash computations was first introduced by \cite{keytrap}. Their work showed that DNSSEC was vulnerable to a new class of attacks that can exhaust CPU resources and thereby achieve Denial of Service on any DNSSEC validating resolver. Their most impactful attack, KeyTrap, achieves a full DoS of DNS resolvers for between 2min and 16h by exploiting colliding key-tags and a large number of signatures, leading to quadratic complexity in validating DNSSEC signatures. Further, their findings include an attack exploiting hash computations over the DS hash that connects a parent zone to a child zone. Specifically, in their attack, they include a large amount of DS hash records in the parent zone and point them to a single entry in the child-zone with a specific key-tag value. Exploiting colliding key-tags, they achieve quadratic complexity in hash computations, requiring the resolver to try each DS record in the parent zone against each DNSKEY in the child-zone. This computational effort allows for a DoS of the resolver. The NSEC3-encloser attack that we study in this work differs significantly in its single-request impact from the attacks described in \cite{keytrap}. Comparing to the KeyTrap attack, the NSEC3-encloser attack inflicts a modest 72x increase in CPU instruction count,\footnote{Measured on Unbound, average over 5 measurements} while KeyTrap increases CPU instructions by a factor of $2\,000\,000$x. Thus, with KeyTrap, a single attacker is able to DoS a resolver for an extended period of time, whereas with NSEC3, a large attack traffic volume is necessary, consisting of hundreds of DNS requests per second to exhaust the CPU of a victim resolver. This is expected, as KeyTrap exploits computationally heavy public key cryptography, while NSEC3 only uses hash calculations, which require less CPU resources. However, while requiring more traffic, the NSEC3 attack can still harm DNS resolvers, as it can create a heavy load on the attacked resolver and therefore lead to substantial degradation of service. 

Our work is also related to downgrade attacks against DNSSEC~\cite{heftrig2023downgrading}. The DNSSEC downgrade attacks however focus on disabling DNSSEC validation but do not have adverse effects on the availability of the victim resolvers.

%% file: 070-conclusions.tex
\section{Conclusions}\label{sec:conclusions}
We perform extensive evaluations of NSEC3-encloser attack and find that it can create a 72x increase in CPU instruction count on victim DNS resolvers. This is much less than the recently disclosed KeyTrap attack, which creates a factor of $2\,000\,000$ increase in CPU instructions count. Our experimental evaluation shows that even the improved implementation of the NSEC3-encloser attack that we developed creates a relatively minor packet loss (between 2.7\% and 30\% depending on the resolver implementation), yet requires a high traffic volume from an adversary and can be easily detected. Therefore we do not expect to see such attacks in the wild. Nevertheless, our study shows that NSEC3-encloser attack points to a potential problem in the resolvers, that was also raised by the NSEC standard specification. In this work, we explore the practical aspects of NSEC across DNS resolver implementations.

We experimentally analyze the role of the different parameters in NSEC3 on the load created on the resolvers and show how to adjust the parameters to optimize the impact of the attack. Although the increase in CPU instruction set is lower than previous attacks on DNS, such as KeyTrap or NRDelegation, using about a hundred packets per second, the adversary can still create a sufficient load on the resolvers, eventually leading to packet loss. The load is created by the iterative application of the hash in NSEC3, and is further exacerbated by the application of salt to the computation of the hash. Multiple hash iterations with salt make zone enumeration attacks more difficult, requiring more resources from the attackers.

Such records can be exploited to exhaust resources on victim resolvers, as we experimentally demonstrate in this work. The effect of resource exhaustion may become even more severe with the new proposal NSEC5 which uses public key operations \cite{cryptoeprint:2017/099}. Our research essentially shows that there is a tradeoff between the privacy and the load on DNS resolvers, which can be exploited for attacks. This tradeoff is also aligned with the question raised by RFC9276: do the increased performance costs justify applying additional hash operations. 

As RFC9276 points out, most of the names published in DNS are typically public and are rarely secret or unpredictable. RFC9276: ``{\em They are published to be memorable, used and consumed by humans. They are often recorded in many other network logs such as email logs, certificate transparency logs, web page links, intrusion-detection systems, malware scanners, email archives, etc. Many times a simple dictionary of commonly used domain names prefixes (www, mail, imap, login, database, etc.) can be used to quickly reveal a large number of labels within a zone.}''

The fundamental question of the tradeoff between privacy of the resources in the DNS zones vs load on the DNS resolvers poses an important decision that the research and operational community need to take.

%% file: main.bbl
\begin{thebibliography}{10}

\bibitem{afek2020nxnsattack}
Yehuda Afek, Anat Bremler-Barr, and Lior Shafir.
\newblock {NXNSAttack}: Recursive {DNS} inefficiencies and vulnerabilities.
\newblock In {\em 29th USENIX Security Symposium (USENIX Security 20)}, pages 631--648. USENIX Association, 2020.

\bibitem{afek2023nrdelegationattack}
Yehuda Afek, Anat Bremler-Barr, and Shani Stajnrod.
\newblock {NRDelegationAttack}: Complexity {DDoS} attack on {DNS} recursive resolvers.
\newblock In {\em 32nd USENIX Security Symposium (USENIX Security 23)}, pages 3187--3204. USENIX Association, 2023.

\bibitem{bau2010security}
Jason Bau and John~C Mitchell.
\newblock A security evaluation of {DNSSEC} with {NSEC3}.
\newblock {\em Cryptology ePrint Archive}, 2010.

\bibitem{bushart2018dns}
Jonas Bushart and Christian Rossow.
\newblock {DNS} unchained: Amplified application-layer {DoS} attacks against {DNS} authoritatives.
\newblock In Michael Bailey, Thorsten Holz, Manolis Stamatogiannakis, and Sotiris Ioannidis, editors, {\em Research in Attacks, Intrusions, and Defenses}, pages 139--160. Springer, 2018.

\bibitem{goldberg2015stretching}
Sharon Goldberg, Moni Naor, Dimitrios Papadopoulos, Leonid Reyzin, Sachin Vasant, and Asaf Ziv.
\newblock Stretching {NSEC3} to the limit: Efficient zone enumeration attacks on {NSEC3} variants.
\newblock Technical report, Boston University, 2015.

\bibitem{gruza_2024_11352869}
Olivia Gruza, Elias Heftrig, Oliver Jacobsen, Haya Schulmann, Niklas Vogel, and Michael Waidner.
\newblock {Goethe-Universitat-Cybersecurity/NSEC3-Encloser- Attack: WOOT'24 Artifact}, May 2024.

\bibitem{keytrap}
Elias Heftrig, Haya Schulmann, Niklas Vogel, and Michael Waidner.
\newblock {The Harder You Try, The Harder You Fail: The KeyTrap Denial-of-Service Algorithmic Complexity Attacks on DNSSEC}.
\newblock In {\em ACM Conference on Computer and Communications Security (CCS)}, 2024.

\bibitem{heftrig2023downgrading}
Elias Heftrig, Haya Shulman, and Michael Waidner.
\newblock {Downgrading DNSSEC: How to Exploit Crypto Agility for Hijacking Signed Zones}.
\newblock In {\em 32nd USENIX Security Symposium (USENIX Security 23)}, pages 7429--7444, 2023.

\bibitem{kumar2013rainbow}
Himanshu Kumar, Sudhanshu Kumar, Remya Joseph, Dhananjay Kumar, Sunil~Kumar Shrinarayan~Singh, Ajay Kumar, and Praveen Kumar.
\newblock Rainbow table to crack password using {MD5} hashing algorithm.
\newblock In {\em 2013 IEEE Conference on Information \& Communication Technologies}, pages 433--439. IEEE, 2013.

\bibitem{mitchelltaking}
Harrison Mitchell.
\newblock Taking the {DNS} for a walk; {NSEC3} prevalence and recoverability.

\bibitem{DBLP:conf/imc/MouraCHWH20}
Giovane C.~M. Moura, Sebastian Castro, Wes Hardaker, Maarten Wullink, and Cristian Hesselman.
\newblock Clouding up the internet: how centralized is {DNS} traffic becoming?
\newblock In {\em Internet Measurement Conference}, pages 42--49. {ACM}, 2020.

\bibitem{cryptoeprint:2017/099}
Dimitrios Papadopoulos, Duane Wessels, Shumon Huque, Moni Naor, Jan Včelák, Leonid Reyzin, and Sharon Goldberg.
\newblock Making {NSEC5} practical for {DNSSEC}.
\newblock Cryptology ePrint Archive, Paper 2017/099, 2017.
\newblock \url{https://eprint.iacr.org/2017/099}.

\end{thebibliography}
